\documentclass[journal=nalefd,manuscript=article]{achemso}

\usepackage[version=3]{mhchem} 

\title{Efficient injection and detection of out-of-plane spins via the anomalous spin Hall effect in permalloy nanowires}


\author{K. S. Das}
\email{k.s.das@rug.nl}
\author{J. Liu}
\author{B. J. van Wees}
\email{b.j.van.wees@rug.nl}
\affiliation{Physics of Nanodevices, Zernike Institute for Advanced Materials, University of Groningen, Nijenborgh 4, 9747 AG Groningen, The Netherlands}
\author{I. J. Vera-Marun}
\email{ivan.veramarun@manchester.ac.uk}
\affiliation{School of Physics and Astronomy, University of Manchester, Manchester M13 9PL, United Kingdom}

\begin{document}

\newpage
\begin{abstract}
We report a novel mechanism for the electrical injection and detection of out-of-plane spin accumulation via the anomalous spin Hall effect (ASHE), where the direction of the spin accumulation can be controlled by manipulating the magnetization of the ferromagnet. This mechanism is distinct from the spin Hall effect (SHE), where the spin accumulation is created along a fixed direction parallel to an interface. We demonstrate this unique property of the ASHE in nanowires made of permalloy (Py), to inject and detect out-of-plane spin accumulation in a magnetic insulator, yttrium iron garnet (YIG). We show that the efficiency for the injection/detection of out-of-plane spins can be up to 50\% of that of in-plane spins. We further report the possibility to detect spin currents parallel to the Py/YIG interface for spins fully oriented in the out-of-plane direction, resulting in a sign reversal of the non-local magnon spin signal. The new mechanisms that we have demonstrated are highly relevant for spin torque devices and applications.
\end{abstract}

\newpage

Electrical injection and detection of spin currents plays an essential role for the technological implementation of spintronics. The conventional way of electrical spin injection is by driving a spin-polarized current from a ferromagnet into a normal metal \cite{jedema_electrical_2001}. 
This method, however, is limited in the scalability and direction of the injected spin current, which is parallel to the charge current, and has motivated the study of alternative methods based on the spin Hall effect (SHE) present in heavy non-magnetic metals \cite{kimura_room-temperature_2007,sinova_spin_2015}. The SHE generates a spin current perpendicular to a charge current, which is particularly significant for spin torque applications \cite{miron_perpendicular_2011,liu_spin-torque_2011,liu_spin-torque_2012,yu_switching_2014,lau_spinorbit_2016} and for spin injection into magnetic insulators \cite{cornelissen_long-distance_2015,kajiwara_transmission_2010,goennenwein_non-local_2015}.

\begin{figure}[h]
	\centering
	\includegraphics[width=0.72\textwidth]{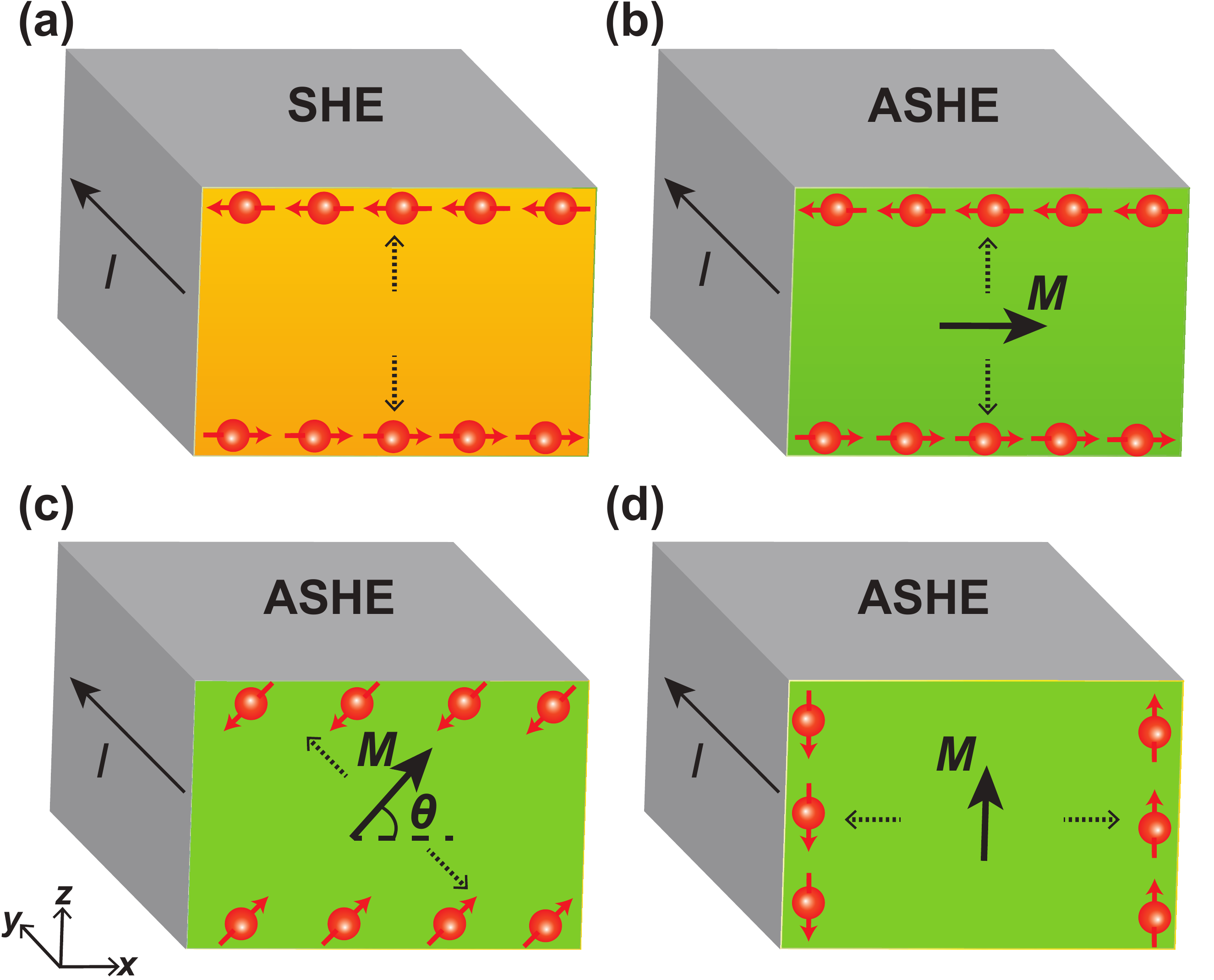}
	\caption{
		Schematic illustration of: \textbf{(a)} the spin Hall effect (SHE) in a metal with high spin-orbit coupling, \textbf{(b-d)} the anomalous spin Hall effect (ASHE) in a ferromagnetic metal for three different orientations of the ferromagnet's magnetization ($\textbf{\textit{M}}$) and a fixed charge current ($\textbf{\textit{I}}$). The magnitude and the direction of the spin current generated due to the ASHE is given by $\textbf{\textit{M}}\times \textbf{\textit{I}}$, with the spin accumulation direction parallel to $\textbf{\textit{M}}$. Spin accumulation with both in-plane and out-of-plane components is generated at the bottom interface when $\textbf{\textit{M}}$ tilts out of the plane, as shown in \textbf{(c)}. The contribution of the out-of-plane component of the spin accumulation at the bottom interface is given by $\sin\theta\cos\theta$ and reaches a maximum of 50\% when $\theta=45^\circ$, compared to the contribution of the in-plane spin component at the bottom interface when $\theta=0^\circ$. Spin accumulation exclusively oriented perpendicular to the top/bottom interface is achieved at the edges when $\textbf{\textit{M}}$ is oriented completely in the out-of-plane direction, as shown in \textbf{(d)}. The dashed arrows indicate the directions of the spin current.      
	}
	\label{fig:Schematic}
\end{figure}

However, the spin direction of the spin accumulation generated via the SHE is fixed, parallel to the interface, depending only on the direction of the charge current through the heavy non-magnetic metal [Fig.~\ref{fig:Schematic}(a)]. Alternatively, the anomalous Hall effect \cite{nagaosa_anomalous_2010} in ferromagnetic metals can be used as a tunable source of transverse spin current, as has been theoretically predicted \cite{taniguchi_spin-transfer_2015,taniguchi_magnetoresistance_2016,taniguchi_magnetoresistance_2017} and recently demonstrated experimentally \cite{das_spin_2017,gibbons_reorientable_2017,qin_nonlocal_2017,iihama_spin-transfer_2018}. We call this phenomenon the anomalous spin Hall effect (ASHE), which generates a spin accumulation oriented parallel to the ferromagnet's magnetization [Fig.~\ref{fig:Schematic}(b) - \ref{fig:Schematic}(d)]. In principle, the ASHE provides a novel way of electrically injecting and detecting a spin accumulation with out-of-plane components, which can be controlled by manipulating the ferromagnet's magnetization.

\begin{figure}[h]
    \centering
        \includegraphics[width=0.8\textwidth]{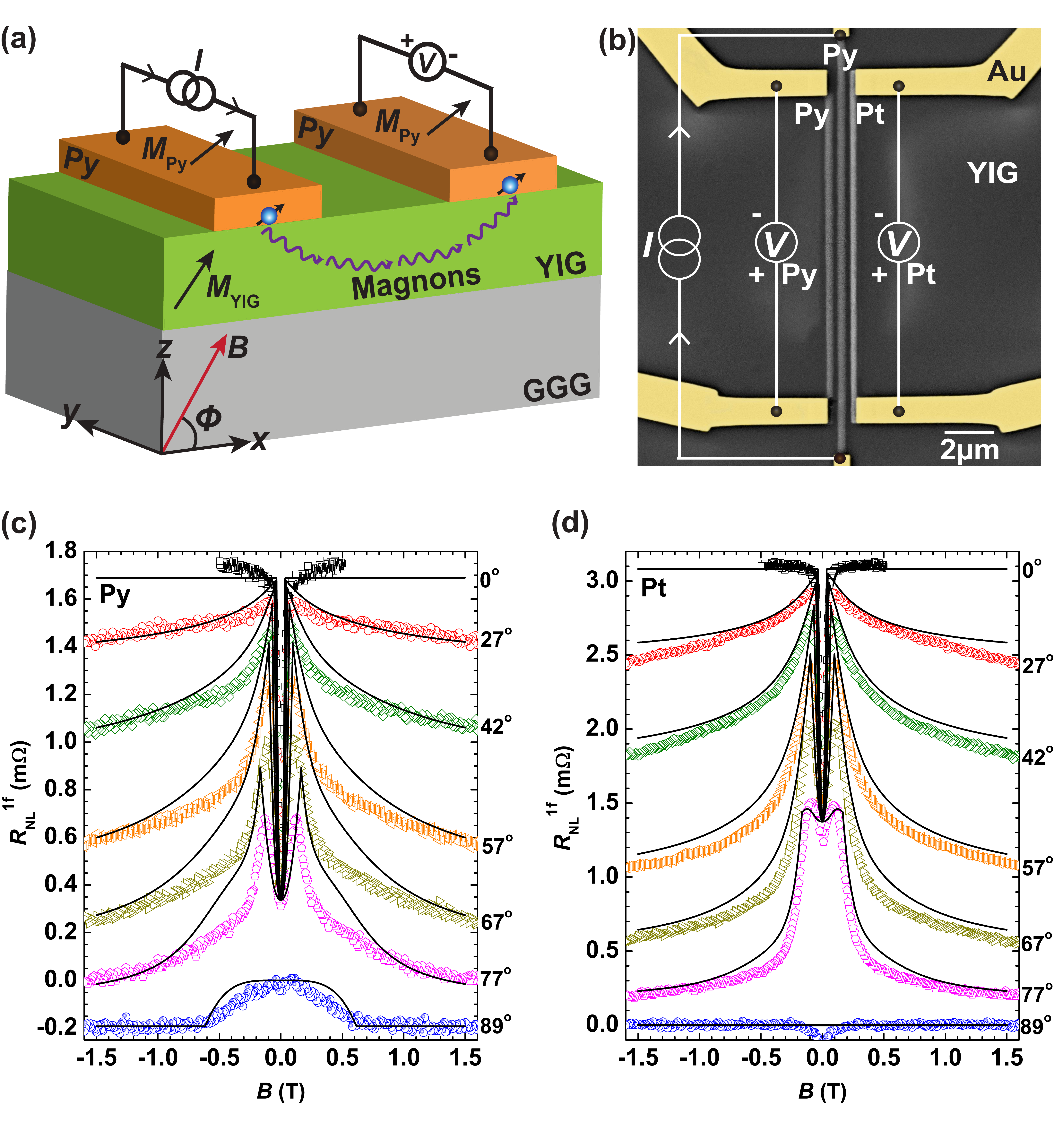}
    \caption{\textbf{(a)} Schematic illustration of the experimental geometry. The ASHE and its reciprocal effect in Py are used to inject and detect magnons in the YIG film. An external magnetic field ($\textbf{\textit{B}}$) is applied in the $xz$-plane, at an angle $\phi$ with respect to the $x$-axis, to manipulate the magnetizations of Py ($\textbf{\textit{M}}_\text{Py}$) and YIG ($\textbf{\textit{M}}_\text{YIG}$). \textbf{(b)} SEM image of a representative device illustrating the electrical connections. An alternating current ($\textbf{\textit{I}}$) is sourced through the injector (middle Py nanowire). The corresponding non-local voltages across the left Py detector ($V_\text{Py}$) and the right Pt detector ($V_\text{Pt}$) are measured simultaneously. \textbf{(c-d)} The first harmonic response of the non-local resistance ($R_\text{NL}^\text{1f}$) is plotted as a function of $\textbf{\textit{B}}$ applied at different angles ($\phi$), measured by the Py detector \textbf{(c)} and the Pt detector \textbf{(d)}. Symbols represent experimental data, while solid black lines are modelled curves following Eq.~\ref{eq:Py_detection} and Eq.~\ref{eq:Pt_detection} for the Py and the Pt detectors, respectively. }
    \label{fig:Experiment}
\end{figure}

Here, we experimentally demonstrate the versatility of the ASHE for electrically injecting and detecting spin accumulation oriented in arbitrary directions, parallel to the ferromagnet's magnetization, in a proof-of-concept device geometry. We utilize the ASHE in a nanowire made of a ferromagnetic metal, permalloy (Ni$_{80}$Fe$_{20}$, Py), to inject a magnon spin accumulation in a magnetic insulator, yttrium iron garnet ($\text{Y}_3\text{Fe}_5\text{O}_{12}$, YIG). The injected magnon spins are electrically detected at a second Py nanowire. 
This non-local geometry, shown in Fig.~\ref{fig:Experiment}(a), and the insulating property of the YIG film ensure that we exclusively address spin-dependent effects, free from magnetoresistance due to the magnetization of the Py nanowire ($\textbf{\textit{M}}_\text{Py}$). 
Moreover, the YIG film serves as a selector of the spin components from the Py injector, since only the spin component parallel to the YIG magnetization ($\textbf{\textit{M}}_\text{YIG}$) will result in the generation of magnon spin accumulation in the YIG film \cite{cornelissen_long-distance_2015}. We apply an external magnetic field ($\textbf{\textit{B}}$) at different out-of-plane angles for the distinct manipulation of the magnetizations $\textbf{\textit{M}}_\text{Py}$ and $\textbf{\textit{M}}_\text{YIG}$. Therefore, we control both the direction of the injected and detected spin accumulation generated by the ASHE (parallel to $\textbf{\textit{M}}_\text{Py}$), and the efficiency of the magnon injection and detection process (via the projection of $\textbf{\textit{M}}_\text{Py}$ on $\textbf{\textit{M}}_\text{YIG}$).
Furthermore, we detect a finite non-local signal with a negative sign when both $\textbf{\textit{M}}_\text{Py}$ and $\textbf{\textit{M}}_\text{YIG}$ are oriented fully perpendicular to the sample ($xy$) plane. We attribute this to a second mechanism of generation and detection of horizontal spin currents, parallel to the Py/YIG interface. The efficiency of this injection/detection mechanism is maximum when the spins are fully oriented in the out-of-plane direction. Besides its possible use for magnon transistor and magnon-based logic operations \cite{chumak_magnon_2014,cornelissen_spin-current-controlled_2018,cramer_magnon_2018}, this model system is also highly relevant for spin torque applications \cite{miron_perpendicular_2011,liu_spin-torque_2011,liu_spin-torque_2012,yu_switching_2014,lau_spinorbit_2016}.

The devices were patterned using electron beam lithography on a 210$\,$nm thick YIG film, grown on a GGG ($\text{Gd}_3\text{Ga}_5\text{O}_{12}$) substrate by liquid-phase epitaxy. A scanning electron microscope (SEM) image of a representative device is shown in Fig.~\ref{fig:Experiment}(b). The devices consist of two Py nanowires (left and middle) and one Pt nanowire (right) with thicknesses of 9$\,$nm (Py) and 7$\,$nm (Pt), respectively. The Py and the Pt nanowires were deposited by d.c.\ sputtering in $\text{Ar}^+$ plasma. Electron beam evaporation was used to deposit the Ti/Au leads and bonding pads following the final lithography step. The middle Py nanowire is used as the spin injector, while the outer Py and Pt nanowires are used as detectors. The width of the middle Py injector is 200$\,$nm and that of the outer Py and Pt detectors is 400$\,$nm. The edge-to-edge distance between the injector and the detectors is 500$\,$nm. 
The electrical connections are also depicted in Fig.~\ref{fig:Experiment}(b). An alternating current ($I$), with an rms amplitude of 310$\,$$\mu$A and frequency of 5.5$\,$Hz, is sourced through the middle Py injector. The non-local voltages across the left Py detector ($V_\text{Py}$) and the right Pt detector ($V_\text{Pt}$) are simultaneously recorded by a phase-sensitive lock-in detection technique. 
The first harmonic response (1$f$) of the non-local voltage corresponds to the linear-regime electrical spin injection and detection via the (A)SHE and their reciprocal processes. 
The second harmonic (2$f$) response, driven by Joule heating at the injector and proportional to $I^2$, corresponds to the thermally generated magnons near the injector via the spin Seebeck effect (SSE) \cite{uchida_spin_2010,cornelissen_long-distance_2015} which travel to the detector. At the Py detector, a lateral temperature gradient along the $x$-axis also contributes to an electrical signal via the anomalous Nernst effect (ANE) \cite{slachter_anomalous_2011,bauer_spin_2012}. The non-local voltage [$V^\text{1(2)f}$] measured across the detectors has been normalized by the injection current ($I$) for the first harmonic response ($R_\text{NL}^\text{1f}=V^\text{1f}/I$) and by $I^2$ for the second harmonic response ($R_\text{NL}^\text{2f}=V^\text{2f}/I^2$). The experiments have been conducted in a low vacuum atmosphere at 293$\,$K.

To explore the injection/detection of out-of-plane spins, we performed magnetic field ($\textbf{\textit{B}}$) sweeps within the $xz$-plane, at different angles $\phi$ with respect to the $x$-axis [see Fig.~\ref{fig:Experiment}(a)]. The first harmonic responses ($R_\text{NL}^\text{1f}$) measured by the Py and the Pt detectors are plotted as a function of $\textbf{\textit{B}}$ in Figs.~\ref{fig:Experiment}(c) and \ref{fig:Experiment}(d), respectively. 
$R_\text{NL}^\text{1f}$ comprises of magnon spin injection and detection due to two different mechanisms: (i) SHE (independent of $\textbf{\textit{M}}_\text{Py}$) and (ii) ASHE (maximum contribution when $\textbf{\textit{M}}_\text{Py}$ is perpendicular to $\textbf{\textit{I}}$) \cite{das_spin_2017}. 
The SHE results in a constant spin accumulation oriented along the $x$-axis at the bottom interface of the injector, which leads to a maximum magnon spin injection when $\textbf{\textit{M}}_\text{YIG}$ is also oriented parallel to the $x$-axis. 
Since the YIG film has a small in-plane coercivity of less than 1$\,$mT, $\textbf{\textit{M}}_\text{YIG}$ will be oriented along the $x$-axis at low magnetic fields. This gives rise to a signal of 0.35$\,$m$\Omega$ at the Py detector [Fig.~\ref{fig:Experiment}(c)] and 1.30$\,$m$\Omega$ at the Pt detector [Fig.~\ref{fig:Experiment}(d)] for $B \sim 0$. At such low fields $\textbf{\textit{M}}_\text{Py}$ is oriented along the Py nanowire ($y$-axis) due to shape anisotropy, thus only the SHE contributes to the magnon injection and detection processes. 
The ASHE starts to contribute when $\textbf{\textit{M}}_\text{Py}$ has a component oriented perpendicular to $\textbf{\textit{I}}$, and becomes maximum when $\textbf{\textit{M}}_\text{Py}$ is parallel to the $x$-axis [see Fig.~\ref{fig:Schematic}(b)]. 
Therefore, the maximum non-local signal is attained for $\phi=0^\text{o}$ when $B > 50$$\,$mT, corresponding to $\textbf{\textit{M}}_\text{Py}$ oriented along the $x$-axis \cite{das_spin_2017}.

As the angle $\phi$ is increased, the $z$-components of $\textbf{\textit{M}}_\text{Py}$ ($M_\text{Py}^z$) and $\textbf{\textit{M}}_\text{YIG}$ ($M_\text{YIG}^z$) increase, while the $x$-components ($M_\text{Py}^x$ and $M_\text{YIG}^x$) decrease. The schematic shown in Fig.~\ref{fig:Schematic}(c) depicts the case when $\textbf{\textit{M}}_\text{Py}$ is oriented at an angle $\theta$ with respect to the positive $x$-axis, such that $0^\circ<\theta<90^\circ$. The contribution of the out-of-plane spin component to the spin accumulation at the bottom interface is given by $\sin\theta\cos\theta$ and reaches a maximum of 50\% when $\theta=45^\circ$, compared to that of the in-plane spin component (given by ${\cos}^2\theta$) when $\theta=0^\circ$. When $\textbf{\textit{M}}_\text{Py}$ is oriented fully perpendicular to the bottom interface [Fig.~\ref{fig:Schematic}(d)], spin accumulation with only out-of-plane components are created at the left and right edges of the Py nanowire. In this case, the spin injection and detection efficiency through the bottom interface is expected to be zero.       

However, when $\textbf{\textit{B}}$ is applied almost perpendicular to the plane of the sample ($\phi=89^\text{o}$) the first harmonic response $R_\text{NL}^\text{1f}$ measured by the Py detector changes sign and becomes negative. This result cannot be explained within the standard framework of (SHE driven) transport dominated by in-plane spins, where a vanishing signal is expected \cite{cornelissen_long-distance_2015,kajiwara_transmission_2010,goennenwein_non-local_2015}. 
We therefore argue that such a negative signal can only be understood by the injection/detection mechanism of spin currents parallel to the $x$-axis via the ASHE, the efficiency of which is maximized for spins oriented fully along the $z$-axis [see Fig.~\ref{fig:Schematic}(d)]. This is consistent with $R_\text{NL}^\text{1f}$ measured by the Pt detector, which is zero, as expected from the lack of the ASHE detection in the Pt nanowire. Thus, a spin accumulation with an exclusively out-of-plane component can only be injected and detected via the ASHE and, in our sample geometry, results in a distinct negative polarity of the non-local signal.

\begin{figure}[h]
	\centering
	\includegraphics[width=0.8\textwidth]{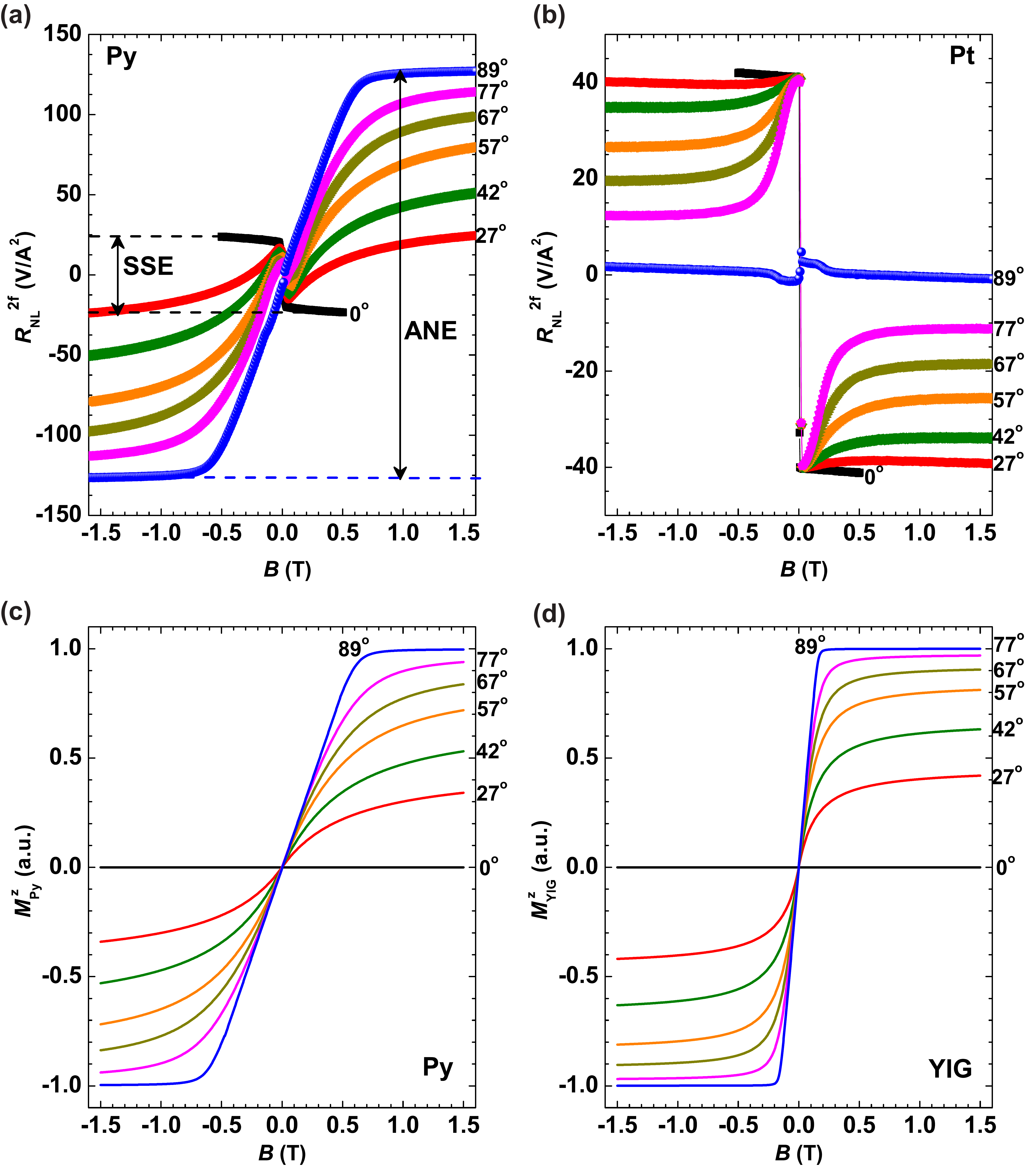}
	\caption{\textbf{(a)} The second harmonic response of the non-local resistance ($R_\text{NL}^\text{2f}$) measured by the Py detector has two contributions: the anomalous Nernst effect (ANE) (proportional to $M_\text{Py}^z$) and the spin Seebeck effect (SSE) (proportional to $M_\text{YIG}^x$). \textbf{(b)} The $R_\text{NL}^\text{2f}$ measured by the Pt detector is only due to the SSE, which decreases as $M_\text{YIG}^z$ increases. $M_\text{Py}^z$ \textbf{(c)} and $M_\text{YIG}^z$ \textbf{(d)} are plotted against $\textbf{\textit{B}}$ for the different out-of-plane angles ($\phi$). The magnetizations are extracted from the Stoner–Wohlfarth model, by fitting the second harmonic responses (discussed in the Supporting Information).}
	\label{fig:SecondHarmonic}
\end{figure}

Further understanding is achieved by studying the second harmonic response measured by the Py and Pt detectors, shown in Figs.~\ref{fig:SecondHarmonic}(a) and \ref{fig:SecondHarmonic}(b), respectively. The temperature gradient generated due to Joule heating at the injector drives the spin Seebeck effect (SSE), and the generated magnons are detected by the Pt nanowire via the inverse spin Hall effect (ISHE) and by the Py detector as a combination of the ISHE and the inverse ASHE. 
In addition to these spin detection processes, at the Py detector the ANE also contributes to $R_\text{NL}^\text{2f}$. Starting with the case $\phi=0^\text{o}$, when $\textbf{\textit{M}}_\text{Py}$ is oriented along the $x$-axis, only the SSE contributes to $R_\text{NL}^\text{2f}$ measured by the Py detector, with a negligible ANE contribution due to the small temperature gradient along the $z$-axis within the Py detector. However, when $\phi \neq 0^\text{o}$ and the $z$-component of $\textbf{\textit{M}}_\text{Py}$ increases, the ANE starts to dominate and is maximized for $\phi=90^\text{o}$, whereas the contribution due to the SSE goes down as the $x$-component of $\textbf{\textit{M}}_\text{YIG}$ decreases. 
We therefore consider ANE $\propto M_\text{Py}^z$ and SSE $\propto M_\text{YIG}^x$, and employ the Stoner-–Wohlfarth model \cite{stoner_mechanism_1948} to extract from $R_\text{NL}^\text{2f}$ the magnetization behaviour of the Py nanowire and the YIG film (see Supporting Information). The extracted $M_\text{Py}^z$ and $M_\text{YIG}^z$ are plotted as a function of $\textbf{\textit{B}}$ for different angles $\phi$ in Figs.~\ref{fig:SecondHarmonic}(c) and \ref{fig:SecondHarmonic}(d), respectively. 

We use the extracted magnetization behaviour of the Py nanowires and the YIG film to model the first harmonic response via the following expressions,

\begin{align}	
R_\text{NL}^\text{1f}\text{(Py)}&=[a M_\text{YIG}^x + b M_\text{Py}^x (\textbf{\textit{M}}_\text{YIG}\cdot \textbf{\textit{M}}_\text{Py})]^2 - [\eta b M_\text{Py}^z (\textbf{\textit{M}}_\text{YIG}\cdot \textbf{\textit{M}}_\text{Py})]^2, \label{eq:Py_detection}\\	    
R_\text{NL}^\text{1f}\text{(Pt)}&=c M_\text{YIG}^x[a M_\text{YIG}^x + b M_\text{Py}^x (\textbf{\textit{M}}_\text{YIG}\cdot \textbf{\textit{M}}_\text{Py})],
\label{eq:Pt_detection}
\end{align}

where, $(\textbf{\textit{M}}_\text{YIG}\cdot \textbf{\textit{M}}_\text{Py})=(M_\text{YIG}^x M_\text{Py}^x + M_\text{YIG}^y M_\text{Py}^y +M_\text{YIG}^z M_\text{Py}^z)$, with $\textbf{\textit{M}}_\text{YIG}$ and $\textbf{\textit{M}}_\text{Py}$ being unitary vectors. The coefficients $a$, $b$ and $c$ can be expressed as 
$a\propto \frac{G_\text{Py}\theta_\text{SH}^\text{Py}\lambda_\text{Py}}{t_\text{Py}\sigma_\text{Py}}$, 
$b\propto \frac{G_\text{Py}\theta_\text{ASH}^\text{Py}\lambda_\text{Py}}{t_\text{Py}\sigma_\text{Py}}$ and 
$c\propto \frac{G_\text{Pt}\theta_\text{SH}^\text{Pt}\lambda_\text{Pt}}{t_\text{Pt}\sigma_\text{Pt}}$ \cite{das_spin_2017}. 
Here, $G_\text{Py(Pt)}$, $\theta_\text{SH}^\text{Py(Pt)}$, $\lambda_\text{Py(Pt)}$, $t_\text{Py(Pt)}$ and $\sigma_\text{Py(Pt)}$ represent the effective spin mixing conductance for the Py(Pt)/YIG interface, the spin Hall angle, the spin relaxation length, the thickness and the charge conductivity of the Py (Pt) nanowire, respectively. $\theta_\text{ASH}^\text{Py}$ is the anomalous spin Hall angle of Py. For the simulations, we use $a=0.58$$\,$$(\text{m}\Omega)^{1/2}$, $b=0.72$$\,$$(\text{m}\Omega)^{1/2}$ and $c=2.37$$\,$$(\text{m}\Omega)^{1/2}$, which are extracted by fitting the experimental data at $\phi=0^\text{o}$, and are in close agreement with our previously reported values \cite{das_spin_2017}. 

\begin{figure}[h]
	\centering
	\includegraphics[width=0.8\textwidth]{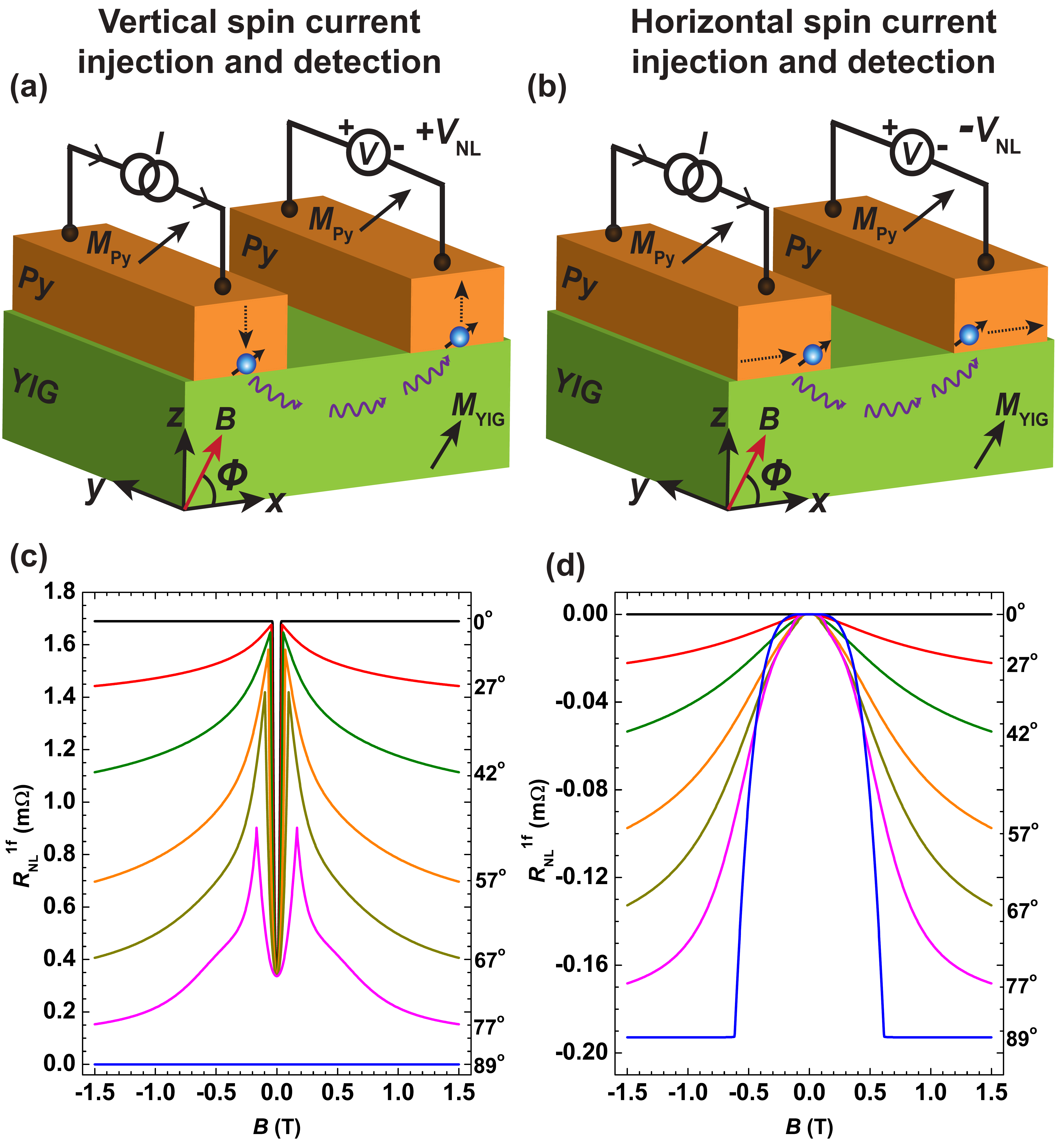}
	\caption{\textbf{(a)} Mechanism for spin current injection and detection along the $-z$ and the $+z$ directions, respectively, resulting in a positive non-local signal ($V_\text{NL}$). This mechanism has the maximum contribution to $V_\text{NL}$ for in-plane spins. \textbf{(b)} Mechanism for spin current injection and the detection along the $x$ direction, parallel to the Py/YIG interface, resulting in a negative $V_\text{NL}$. This mechanism has the maximum contribution to $V_\text{NL}$ for out-of-plane spins. The individual contribution of the two different mechanisms to the non-local resistance ($R_\text{NL}$) measured by the Py detector, following Eq.~\ref{eq:Py_detection}, has been plotted in \textbf{(c)} for the injection and detection of vertical spin current, and in \textbf{(d)} for the injection and detection of horizontal spin current.}
	\label{fig:Model}
\end{figure}

The first part of Eq.~\ref{eq:Py_detection} within the square brackets accounts for the spin current directed perpendicular to the Py/YIG interface, as depicted in Fig.~\ref{fig:Model}(a). The term with the coefficient $a$ is related to the (constant) spin accumulation along the $x$-axis due to the SHE in Py, which is independent of $\textbf{\textit{M}}_\text{Py}$. This term only depends on $M_\text{YIG}^x$ since the generation of magnons is proportional to the projection of $\textbf{\textit{M}}_\text{YIG}$ on the spin accumulation direction. The term with the coefficient $b$ is related to the ASHE in Py, which is maximized when $\textbf{\textit{M}}_\text{Py}$ is parallel to the $x$-axis. The ASHE generates a spin accumulation parallel to $\textbf{\textit{M}}_\text{Py}$, thus the magnon generation is also proportional to the projection of $\textbf{\textit{M}}_\text{Py}$ on $\textbf{\textit{M}}_\text{YIG}$. This term includes both the in-plane and the out-of-plane components of the spin accumulation. Since the injection and detection processes are reciprocal, the term within the square brackets is squared.

The second part of Eq.~\ref{eq:Py_detection}, within the square brackets and preceded by a negative sign, accounts for the spin current parallel to the Py/YIG interface, as depicted in Fig.~\ref{fig:Model}(b). The contribution of this part to the magnon injection and detection processes is maximum for out-of-plane spins. It is clear from the symmetry of the ASHE and our measurement geometry that the detection of such in-plane spin currents, with spins oriented in the out-of-plane direction, will result in a negative non-local signal measured by the Py detector [Figs.~\ref{fig:Model}(a) and \ref{fig:Model}(b)]. Moreover, the parameter $\eta$ tells us the efficiency of detecting spin currents parallel to the interface for out-of-plane spins as compared to that of detecting spin currents perpendicular to the interface for in-plane spins. By fitting the experimental data, we obtain $\eta=61\%$. Note that the detection of the spin current parallel to the interface is achieved exclusively via the ASHE. This is evident in the lack of a negative signal while using the Pt nanowire as a detector, where the only detection mechanism is via the ISHE. Thus the Pt nanowire is only sensitive to the spin current perpendicular to the Pt/YIG interface for in-plane spins. Eq.~\ref{eq:Pt_detection} describes the spin injection by the Py injector (following Eq.~\ref{eq:Py_detection}) and the detection via the ISHE in the Pt nanowire.    

The simulated curves, following Eq.~\ref{eq:Py_detection} and Eq.~\ref{eq:Pt_detection}, are shown as solid black lines in Figs.~\ref{fig:Experiment}(c) and \ref{fig:Experiment}(d), respectively. This modelling for all tilt angles ($\phi$) employs the same values for the parameters $a$, $b$ and $c$ as those extracted from the in-plane measurements at $\phi = 0^\circ$. The satisfactory agreement with the experimental data, both in terms of magnitude and lineshape,  demonstrates that our model captures the dominant physics of the out-of-plane spin injection and detection processes. To achieve further insight, we separate the modelled contributions of the spin current perpendicular to the interface and the spin current parallel to the interface to the non-local signal at the Py detector, following Eq.~\ref{eq:Py_detection}. The results, shown in Figs.~\ref{fig:Model}(c) and \ref{fig:Model}(d), present in an explicit manner the contribution of the two different mechanisms of detecting the spin current oriented along the $z$-axis and that along the $x$-axis, respectively, with increasing $\phi$. 

Note that, although we understand the different symmetries of the injection/detection mechanisms depicted in Figs.~\ref{fig:Model}(a) and \ref{fig:Model}(b), we do not fully understand why these mechanisms have comparable efficiencies, given the specific cross sections of the nanowires. The minor disagreement between model and experiment, observed at intermediate values of $\textbf{\textit{B}}$, can be attributed either to the extraction method of the magnetization behaviour of the Py nanowires and the YIG film, shown in Figs.~\ref{fig:SecondHarmonic}(c) and \ref{fig:SecondHarmonic}(d), or could hint to a subtle effect not present in our description. To explore the latter, we have considered a second set of fitting curves with a non-constant $b$ parameter, motivated by recent studies on spin rotation symmetry and dephasing \cite{humphries_observation_2017}. The apparent variation of the spin injection and detection processes due to a tilted $\textbf{\textit{M}}_\text{Py}$ is of only up to $\sim 20 \%$ (see Supporting Information). Finally, control experiments with a different architecture using a Pt injector and a Pt detector have been performed, confirming the absence of  injection and detection of out-of-plane spins via the SHE alone (see Supporting Information).

The present demonstration of electrical injection and detection of spin accumulation in arbitrary directions is highly desirable in spintronics. We envision that the use of out-of-plane spins within transverse spin currents, in a common ferromagnetic metal like permalloy, has the potential to impact spintronic-based technologies like spin-transfer-torque memories \cite{miron_perpendicular_2011,liu_spin-torque_2011,liu_spin-torque_2012,yu_switching_2014,lau_spinorbit_2016} and logic devices \cite{chumak_magnon_2014,cornelissen_spin-current-controlled_2018,cramer_magnon_2018}. Further remains both on the fundamental understanding, and on the possible implications for previous SHE studies, where the control of spin transport efficiency and directionality enabled by the ASHE \cite{das_spin_2017,gibbons_reorientable_2017,qin_nonlocal_2017,iihama_spin-transfer_2018} has not been hitherto considered.


\begin{acknowledgement}
	We thank J.\ G.\ Holstein, H.\ M.\ de Roosz, H.\ Adema and T.\ Schouten for their technical assistance. We acknowledge the financial support of the Zernike Institute for Advanced Materials, the Future and Emerging Technologies (FET) programme within the Seventh Framework Programme for Research of the European Commission, under FET-Open Grant No.~618083 (CNTQC) and the research program Magnon Spintronics (MSP) No. 159, financed by the Nederlandse Organisatie voor Wetenschappelijk Onderzoek (NWO). This project is also financed by the NWO Spinoza prize awarded to Prof. B. J. van Wees by the NWO.
\end{acknowledgement}

\begin{suppinfo}

The Supporting Information contains details on determination the Py and the YIG magnetization orientations using the Stoner--Wohlfarth model. The Supporting Information also contains the modelling results of the first harmonic non-local resistance with an angle-dependent $b$-parameter. The experimental results on a control device with a Pt injector and a Pt detector have also been included in the Supporting Information.
\end{suppinfo}


\providecommand{\latin}[1]{#1}
\makeatletter
\providecommand{\doi}
{\begingroup\let\do\@makeother\dospecials
	\catcode`\{=1 \catcode`\}=2 \doi@aux}
\providecommand{\doi@aux}[1]{\endgroup\texttt{#1}}
\makeatother
\providecommand*\mcitethebibliography{\thebibliography}
\csname @ifundefined\endcsname{endmcitethebibliography}
{\let\endmcitethebibliography\endthebibliography}{}

\newpage

\appendix

\section{Efficient injection and detection of out-of-plane spins via the anomalous spin Hall effect in permalloy nanowires: Supporting Information}

\subsection{Determination of the Py and the YIG magnetization orientations}

In this section, we will discuss the procedure for determining the orientation of $M_\text{Py}$ and $M_\text{YIG}$.

\begin{figure}[tbp]
	\includegraphics[width=0.9\linewidth]{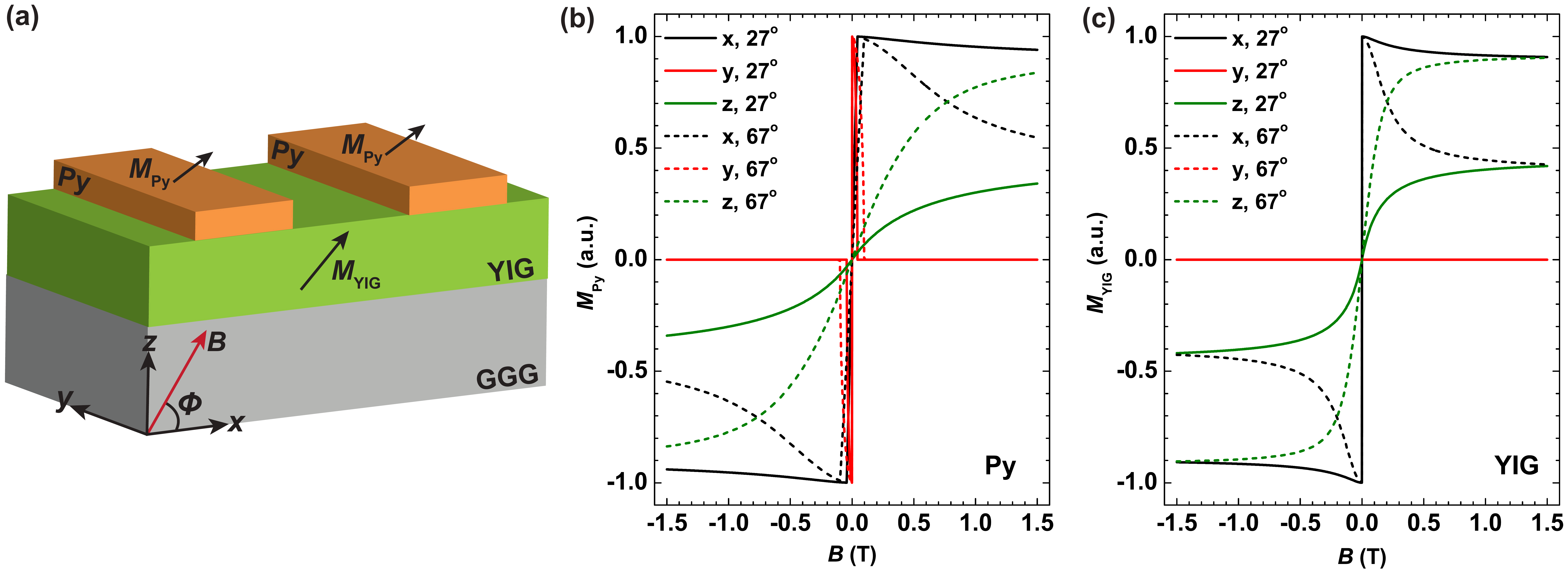} 
	\caption{\textbf{(a)} Schematic illustration of the measured device in a $xyz$-coordinate system. The YIG film lies in the $xy$-plane. The Py nanowires are aligned along the $y$-axis. An external magnetic field $\textit{\textbf{B}}$ is applied in the $xz$-plane at an angle of $\phi$ with respect to the $x$-axis. Modelled results for the normalized magnetization components along $x$-, $y$- and $z$-axes as a function of magnetic field $\textbf{\textit{B}}$ for \textbf{(b)} Py (${\textit{M}}_{\textrm{Py}}^{\textrm{x}}$, ${\textit{M}}_{\textrm{Py}}^{\textrm{y}}$ and ${\textit{M}}_{\textrm{Py}}^{\textrm{z}}$) and \textbf{(c)} YIG (${\textit{M}}_{\textrm{YIG}}^{\textrm{x}}$, ${\textit{M}}_{\textrm{YIG}}^{\textrm{y}}$ and ${\textit{M}}_{\textrm{YIG}}^{\textrm{z}}$). Here, we show the examples under the condition of $\phi= 27\,^{\circ}$ (solid lines), $67\,^{\circ}$ (dashed lines). For Py, we use the following parameters: $\textit{M}_{\textrm{Py}}^{\textrm{s}}=5.2\times10^{5}\,$$\,$[A/m], $\textit{K}_{\textrm{Py}}^{\textrm{z}}=160\,$$\,$[kJ/m$^{3}$], $\textit{K}_{\textrm{Py}}^{\textrm{y}}\approx0\,$$\,$[kJ/m$^{3}$] and $\textit{K}_{\textrm{Py}}^{\textrm{x}}=10\,$$\,$[kJ/m$^{3}$]. For YIG, we use: $\textit{M}_{\textrm{YIG}}^{\textrm{s}}=2.1\times10^{5}\,$$\,$[A/m], $\textit{K}_{\textrm{YIG}}^{\textrm{z}}=17\,$$\,$[kJ/m$^{3}$] and $\textit{K}_{\textrm{YIG}}^{\textrm{y}}=\textit{K}_{\textrm{YIG}}^{\textrm{x}}=1\,$$\,$[kJ/m$^{3}$].}
	\label{figS1}
\end{figure}

A magnetic field $\textit{\textbf{B}}$ is applied at an angle $\phi$ with respect to the $x$-axis, as shown in Fig.~\ref{figS1}(a). For every field-sweep curve, we sweep the magnitude of $\textit{\textbf{B}}$ in both positive and negative directions. The applied $\textit{\textbf{B}}$ has both in-plane and out-of-plane components with respect to the plane of the YIG film. The sample is aligned in such a way that $\textit{\textbf{B}}$ is in the $xz$-plane. Therefore, we can simply express $\textit{\textbf{B}}$ as
\begin{equation}
\textit{\textbf{B}}=(B^x,0,B^z)
=(B\cos\phi,0,B\sin\phi)
\end{equation}
where $\phi$ is the angle between $\textit{\textbf{B}}$ and x-axis. The orientation of $\textit{\textbf{M}}_{\textrm{Py}}$ and $\textit{\textbf{M}}_{\textrm{YIG}}$ does not only depend on $\textit{\textbf{B}}$ but also the saturation magnetization ($\textit{M}_{\textrm{Py}}^{\textrm{s}}$ and $\textit{M}_{\textrm{YIG}}^{\textrm{s}}$) and the shape of the magnets.

Firstly, we write the magnetization of the Py as

\begin{equation}
\textit{\textbf{M}}_{\textrm{Py}}=(M^{\textrm{x}}_{\textrm{Py}},M^{\textrm{y}}_{\textrm{Py}},M^{\textrm{z}}_{\textrm{Py}}),
\end{equation}

and we define the angle between $\textit{\textbf{M}}_{\textrm{Py}}$ and three coordinate axes, $\theta_{i}$, as 
\begin{equation}
\cos\theta_{\textrm{i}}=\frac{M^{\textrm{i}}_{\textrm{Py}}}{M^{\textrm{s}}_{\textrm{Py}}},
\end{equation}
where $\textrm{i}=x,y,z$.

In order to find out $\textbf{\textit{M}}_{\textrm{Py}}$ under a given $\textit{\textbf{B}}$, we can write down the magnetism-related energy density $\varepsilon_{\textrm{m}}$ for Py

\begin{equation}
\varepsilon^{\textrm{m}}_{\textrm{Py}}= E^{\textrm{Zeeman}}_{\textrm{Py}}+E^{\textrm{ani}}_{\textrm{Py}}
\end{equation}
where the first term is the Zeeman energy term 

\begin{equation}
E^{\textrm{Zeeman}}_{\textrm{Py}}=-\textbf{\textit{M}}_{\textrm{Py}}\cdot\textbf{\textit{B}},
\end{equation}
and the second term is the anisotropy term
\begin{equation}
E^{\textrm{ani}}_{\textrm{Py}}=\sum_{\textrm{i}=x,y,z} K_{\textrm{Py}}^{\textrm{i}}\sin^2\theta_{\textrm{i}},
\end{equation}
where $K_{\textrm{Py}}^{\textrm{i}}$ is the anisotropy constant of Py along three axes. Due to the shape of Py bar, $\textbf{\textit{M}}_{\textrm{Py}}$ mostly like to align in the plane of the film and along the bar, i.e. $y$-axis. This translates to a relation of $K_{\textrm{Py}}^{z}>K_{\textrm{Py}}^{x}>K_{\textrm{Py}}^{y}$. To determine the orientation of $\textbf{\textit{M}}_{\textrm{Py}}$, we can find out the energy minimum by $\partial\varepsilon^{\textrm{m}}/\partial\theta_{\textrm{i}}=0$ and $\partial^{2}\varepsilon^{\textrm{m}}/\partial^{2}\theta_{\textrm{i}}>0$. When $\varepsilon^{\textrm{m}}$ reaches its minimum, we obtain 
\begin{equation}
\cos\theta_{\textrm{x}}=\frac{M_{\textrm{Py}}B^{\textrm{x}}}{2\,(K_{\textrm{Py}}^{\textrm{y}}-K_{\textrm{Py}}^{\textrm{x}})},
\label{cos_theta_x}
\end{equation}
\begin{equation}
\cos\theta_{\textrm{z}}=\frac{M_{\textrm{Py}}B^{\textrm{z}}}{2\,(K_{\textrm{Py}}^{\textrm{y}}-K_{\textrm{Py}}^{\textrm{z}})},
\label{cos_theta_z}
\end{equation}
\begin{equation}
\cos^{2}\theta_{\textrm{y}}=1-\cos^{2}\theta_{\textrm{x}}-\cos^{2}\theta_{\textrm{z}},
\label{cos_theta_y}
\end{equation}
which tells us the orientation of $\textbf{\textit{M}}_{\textrm{Py}}$. Eqs. \ref{cos_theta_x}, \ref{cos_theta_z} and \ref{cos_theta_y} hold with increasing the field $B$ until it reaches the critical magnetic field $B_{\textrm{c}}$, where $\cos^{2}\theta_{\textrm{y}}=0$.  For $B>B_{\textrm{c}}$, the magnetization of Py lies in the $xz$-plane, namely $\textit{M}^{\textrm{y}}_{\textrm{Py}}=0$. Larger tilting angle $\phi$ corresponds to larger $B_{\textrm{c}}$. In the regime where $B>B_{\textrm{c}}$, the magnetization of Py becomes $\textit{\textbf{M}}_{\textrm{Py}}=(M_{\textrm{Py}}^{\textrm{x}},0,M_{\textrm{Py}}^{\textrm{z}})$ and $\cos\theta_{\textrm{y}}=0$. Doing the same procedure of finding the minimum of $\varepsilon^{\textrm{m}}$ for Py, we can obtain the relation of $\theta_{\textrm{x}}$ as
\begin{equation}
M_{\textrm{Py}}^{\textrm{s}}\,B^{\textrm{x}}\sin\theta_{\textrm{x}}-M_{\textrm{Py}}^{\textrm{s}}\,B^{\textrm{z}}\cos\theta_{\textrm{x}}=2\,(K_{\textrm{Py}}^{\textrm{z}}-K_{\textrm{Py}}^{\textrm{x}})\,\sin\theta_{\textrm{x}}\cos\theta_{\textrm{x}},
\label{cos_theta_x_largeB}
\end{equation}
from which we can model the magnetization behaviour of Py in the large field regime ($B>B_{\textrm{c}}$). Combined with the solution of Eqs. \ref{cos_theta_x}, \ref{cos_theta_z} and \ref{cos_theta_y} for $B<B_{\textrm{c}}$, we get the behaviour of $\textbf{\textit{M}}_{\textrm{Py}}$ in the full field range. Here, we give two examples of the modelled $\textbf{\textit{M}}_{\textrm{Py}}$ for $\phi= 27\,^{\circ}, 67\,^{\circ}$ as shown in Fig.~$\ $\ref{figS1}(b).

Secondly, the orientation of  $\textit{\textbf{M}}_{\textrm{YIG}}$ is also related to $\textit{\textbf{B}}$, the saturation magnetization ($\textit{M}_{\textrm{YIG}}^{\textrm{s}}$) and the shape of the magnet, i.e. the 210-nm-thick thin YIG film in our case. We define the angle between $\textit{\textbf{M}}_{\textrm{YIG}}$ and three coordinate axes, $\gamma_{i}$, as 

\begin{equation}
\cos\gamma_{\textrm{i}}=\frac{M^{\textrm{i}}_{\textrm{YIG}}}{M^{\textrm{s}}_{\textrm{YIG}}}
\end{equation}
where $\textrm{i}=x,y,z$. Since the YIG film is a very soft magnet with significant in-plane shape anisotropy, i.e. the in-plane coercive field is $\sim0.6\,$$\,$mT and the out-of-plane coercive field is $\sim200\,$$\,$mT. We assume it has an isotropic easy plane and an out-of-plane hard axis. We also assume that $\textit{\textbf{M}}_{\textrm{YIG}}$ lies in the same plane as $\textit{\textbf{B}}$, i.e. $xz$-plane. Therefore, we can write down the $\textit{\textbf{M}}_{\textrm{YIG}}$ as

\begin{equation}
\textit{\textbf{M}}_{\textrm{YIG}}=(M^{\textrm{x}}_{\textrm{YIG}},0,M^{\textrm{z}}_{\textrm{YIG}}),
\end{equation}
which gives rise to the similar situation for the Py magnetization when $B>B_{\textrm{c}}$. In order to find out the orientation of the magnetization under an external magnetic field $\textit{\textbf{B}}$, we can write down the magnetism-related energy density $\varepsilon^{\textrm{m}}_{\textrm{YIG}}$ for YIG. Doing the minimizing procedure for $\varepsilon^{\textrm{m}}_{\textrm{YIG}}$, we obtain the formula with the same form as Eq.~\ref{cos_theta_x_largeB} for YIG as

\begin{equation}
M_{\textrm{YIG}}^{\textrm{s}}\,B^{\textrm{x}}\sin\gamma_{\textrm{x}}-M_{\textrm{YIG}}^{\textrm{s}}\,B^{\textrm{z}}\cos\gamma_{\textrm{x}}=2\,(K_{\textrm{YIG}}^{\textrm{z}}-K_{\textrm{YIG}}^{\textrm{x}})\,\sin\gamma_{\textrm{x}}\cos\gamma_{\textrm{x}},
\label{cos_theta_x_largeB_YIG}
\end{equation}
based on which we model the magnetization behaviour of YIG. In Fig.~$\ $\ref{figS1}(c), we give two examples of the modelled $\textbf{\textit{M}}_{\textrm{YIG}}$ for $\phi= 27\,^{\circ}, 67\,^{\circ}$. 

\begin{figure}[tbp]
	\includegraphics[width=0.9\linewidth]{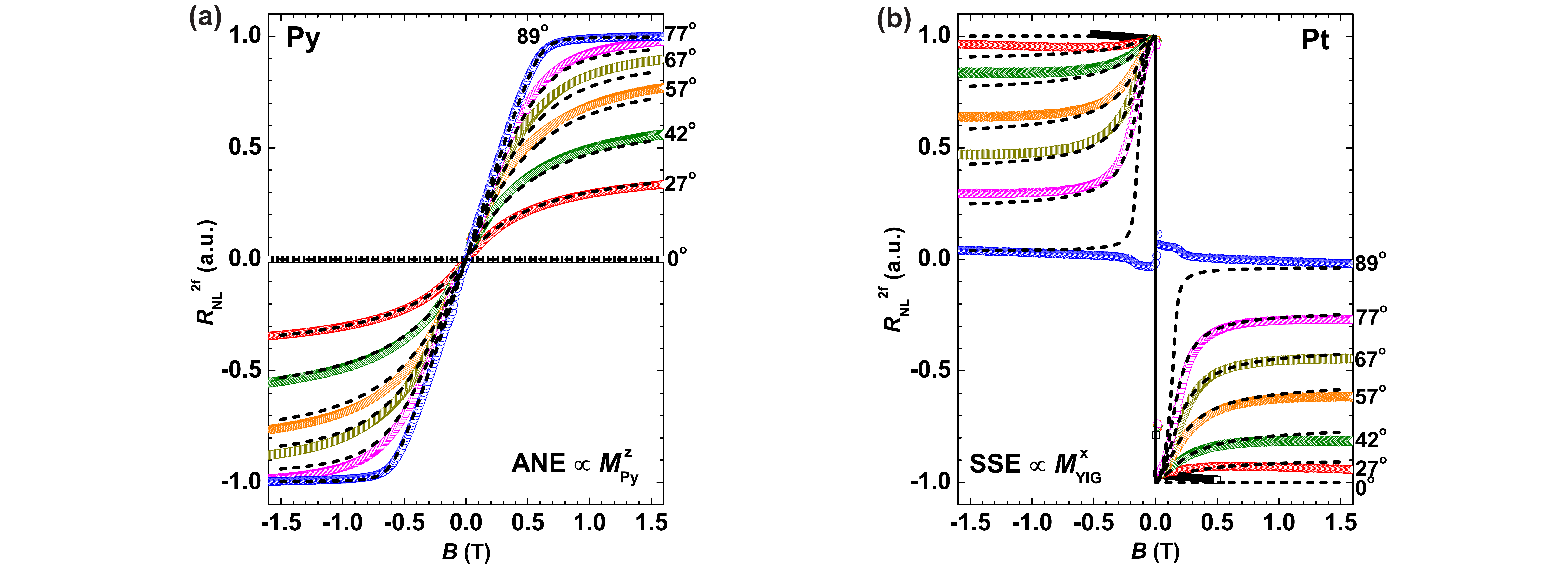} 
	\caption{Normalized second harmonic non-local resistance as a function of the magnetic field strength $B$ with different tilting angles $\phi$ for \textbf{(a)} the Py and \textbf{(b)} the Pt detector. In \textbf{(a)}, we have subtracted the spin Seebeck contribution from the measured signals and we attribute the field-dependent response shown here to only the anomalous Nernst effect of the Py detector, which scales with $\textit{M}_{\textrm{Py}}^{\textrm{z}}$. In \textbf{(b)}, where the detector is a Pt nanowire, the field-dependent behaviour is only due to the spin Seebeck effect of YIG, which is proportional to the $\textit{M}_{\textrm{YIG}}^{\textrm{x}}$. The symbols represent the measured second harmonic data, while the dashed black lines represent the modelled results based on the magnetization behaviour of the Py nanowire and the YIG film, as shown in Fig.$\ $\ref{figS1}.}
	\label{figS2}
\end{figure}

We compare the modelled $\textit{\textbf{M}}_{\textrm{Py}}$ and $\textit{\textbf{M}}_{\textrm{YIG}}$ with the measured second harmonic non-local resistance detected by the Py and the Pt detectors, as shown in Fig.~\ref{figS2}. Note that, we have not considered any interfacial exchange interaction between the Py nanowires and the YIG thin film in modelling the magnetization behaviours. The excellent agreement between the experimental data and the modelled magnetization curves with values of saturation magnetization close to that reported in literature, confirms the absence of any significant interfacial exchange interaction.

\subsection{Modelling the first harmonic non-local resistance with an angle-dependent $b$-parameter}

\begin{figure}[tbp]
	\includegraphics[width=0.9\linewidth]{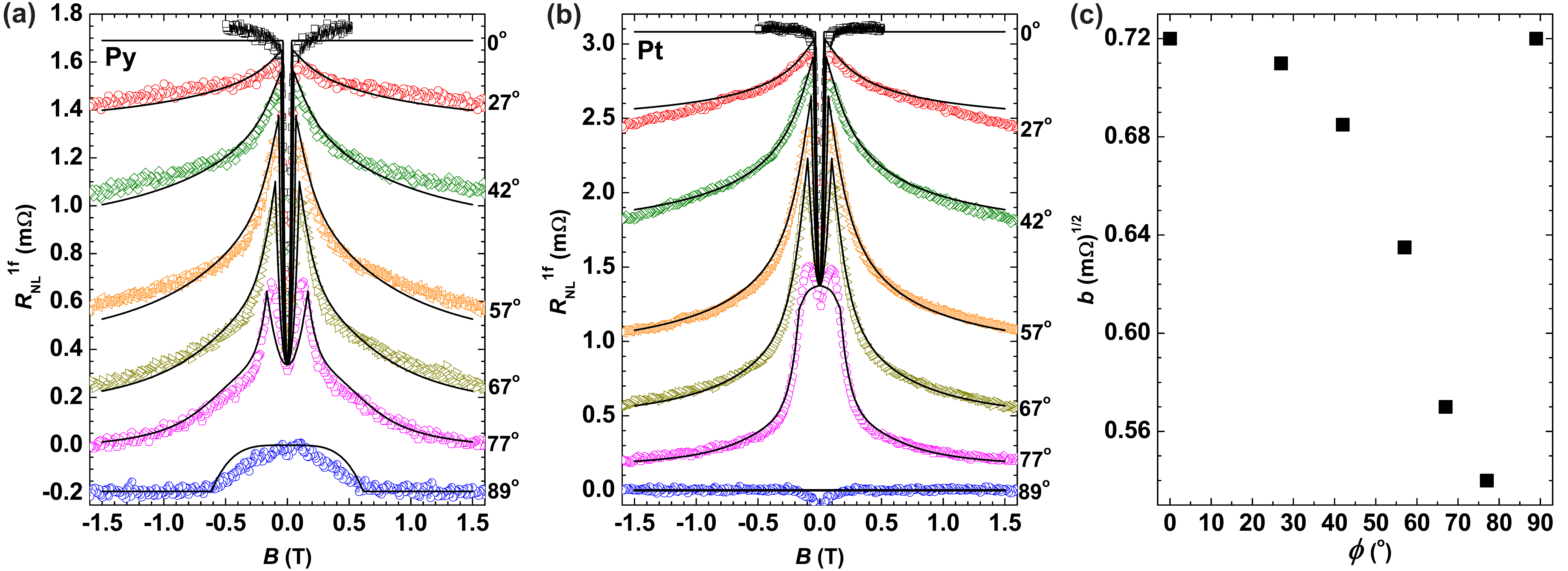} 
	\caption{The first harmonic response of the non-local resistance ($R_\text{NL}^\text{1f}$) is plotted as a function of $\textbf{\textit{B}}$ applied at different angles ($\phi$) in the $xz$-plane, measured by the Py detector \textbf{(a)} and the Pt detector \textbf{(b)}. The symbols represent the experimental data, while the solid black lines represent the modelled curves considering a dependence of the parameter $b$ on $\phi$, as shown in \textbf{(c)}.}
	\label{figS3}
\end{figure}

As discussed in the main text, we obtain a satisfactory agreement between our modelled results (following Eqs.~1 and 2 of the main text) and the experimental data, successfully capturing the physics behind the injection and detection of spin accumulation with out-of-plane components. The modelled results in the main text consider a fixed $b$-parameter, obtained by fitting the in-plane magnetic field sweep data ($\phi=0^{\circ}$). Here, we present another set of modelling results by considering a dependence of the $b$-parameter on $\phi$. The modelled curves, along with the experimental data, are shown in Figs.~\ref{figS3}(a) and \ref{figS3}(b) for the Py detector and the Pt detector, respectively. We obtain an excellent agreement between the model and the measured data at all magnetic field regimes, both for the Py and the Pt detectors. Here, we have used $b$ as a fitting parameter for obtaining the modelled results. Interestingly, we observe a systematic decrease in $b$ from 0.72$\,$$(\text{m}\Omega)^{1/2}$ for $\phi=0^{\circ}$ to 0.54$\,$$(\text{m}\Omega)^{1/2}$ for $\phi=77^{\circ}$. However, for $\phi=89^{\circ}$, the value of $b$ again increases to 0.72$\,$$(\text{m}\Omega)^{1/2}$. At this point we are unsure about the exact physical origin of this behaviour and further study needs to be performed to pinpoint the actual reason. However, this does not contradict our observations and the model presented in the main text, which accurately reproduces the lineshape and the magnitude of the experimentally obtained results in the low and high magnetic field regimes. 

\subsection{Control device with Pt injector and Pt detector}

\begin{figure}[b]
	\includegraphics[width=0.9\linewidth]{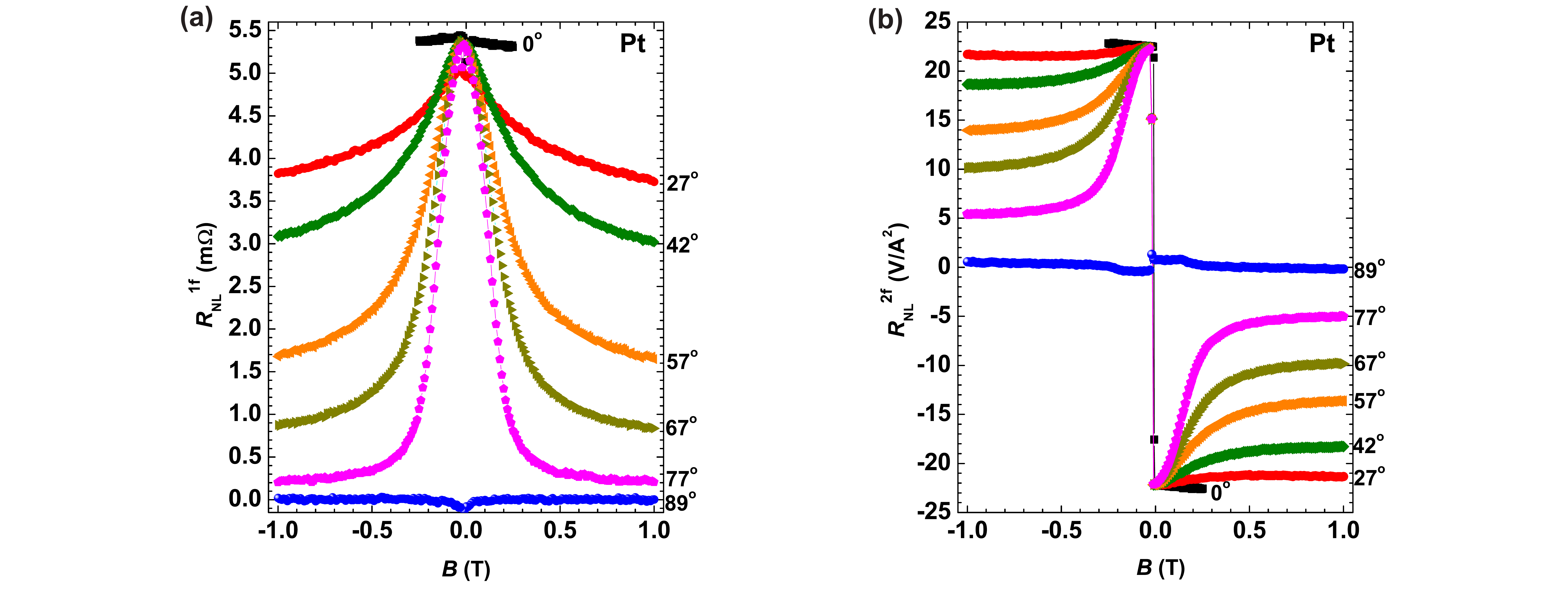} 
	\caption{\textbf{(a)} The first ($R_\text{NL}^\text{1f}$) and \textbf{(b)} the second ($R_\text{NL}^\text{2f}$) harmonic responses of the non-local signal, plotted as a function of $\textbf{\textit{B}}$ applied at different angles ($\phi$) in the $xz$-plane, measured by the Pt detector. This data corresponds to a control device with a Pt injector and a Pt detector.}
	\label{figS4}
\end{figure}

As a control experiment, we fabricated a device with a Pt injector and a Pt detector on the same chip and with the same edge-to-edge separation of 500$\,$nm. Magnetic field sweep measurements were carried out at different tilt angles ($\phi$) in the $xz$-plane, as described in the main text. The first ($R_\text{NL}^\text{1f}$) and the second ($R_\text{NL}^\text{2f}$) harmonic responses of the non-local signal, measured by the Pt detector, are shown in Figs.~\ref{figS4}(a) and \ref{figS4}(b), respectively. Fig.~\ref{figS4}(b) shows that the magnetization behaviour of the YIG film for this control Pt-Pt device is same as that of the Py devices. More importantly, Fig.~\ref{figS4}(a) shows the absence of the modulation of the non-local signal via the ASHE, as opposed to the Py-Py and the Py-Pt devices (discussed in the main text) and the sign reversal at $\phi=89^\circ$ corresponding to the injection and detection of spins oriented fully perpendicular to the Pt/YIG interface. Thus, this control experiment demonstrates that the out-of-plane spin injection and detection achieved via the ASHE, is absent in the pure SHE case.     

\end{document}